\documentclass[twocolumn]{aastex631}

\usepackage{amsmath} 
\usepackage{subfigure}
\usepackage{float}
\usepackage{comment}
\shorttitle{Rapid growth of SMBHs}
\shortauthors{Lin, Chen and Hwang}
\graphicspath{{./}{figures/}}
\begin{document}

% Template \aastex Article with Examples
\title{Rapid Growth of Galactic Supermassive Black Holes through Accreting Giant Molecular Clouds during Major Mergers of their Host Galaxies}

\author[0000-0002-9885-587X]{Chi-Hong\ Lin}
\affiliation{Institute of Astronomy and Astrophysics, Academia Sinica, Taipei 10617, Taiwan R.O.C.}

\author[0000-0002-4848-5508]{Ke-Jung\ Chen}

\affiliation{Institute of Astronomy and Astrophysics, Academia Sinica, Taipei 10617, Taiwan R.O.C.}

%\collaboration{6}{(AAS Journals Data Editors)}

\author[0000-0002-3658-0903]{Chorng-Yuan\ Hwang}
\affiliation{Institute of Astronomy, National Central University, Taoyuan 32001, Taiwan R.O.C.} 

%% automatically takes care of all commas and "and "s between authors names.


%% Mark off the abstract in the "abstract" environment. 

%-------------------- %%% Section 0 Abstract %%% -------------------%

\begin{abstract}

Understanding the formation of the supermassive black holes (SMBHs) present in the centers of galaxies is a crucial topic in modern astrophysics. Observations have detected the SMBHs with mass $M$ of $10^{9}\, \rm M_\odot$ in the high-redshift galaxies with $\rm z\sim7$. However, how SMBHs grew to such huge masses within the first billion years after the big bang remains elusive. One possible explanation is that SMBHs grow quickly through the frequent mergers of galaxies, which provides sustainable gas to maintain rapid growth. This study presents the hydrodynamics simulations of the SMBHs' growth with their host galaxies using the \texttt{GIZMO} code. In contrast to previous simulations, we have developed a giant molecular cloud (GMC) model by separating molecular-gas particles from the atomic-gas particles and then evolving them independently. During major mergers, we show that the more massive molecular gas particles cloud bear stronger dynamical friction. Consequently, GMCs are substantially accreted onto the galactic centers that grow SMBHs from $\sim 10^{7}$\, $\rm M_\odot$ to $\sim 10^{9}\, \rm M_\odot$ within 300 Myr, explaining the rapid growth of SMBHs, and this accretion also triggers a violent starburst at the galactic center. Furthermore, we examine the impact of minor mergers on the bulge of a Milky-Way-like galaxy and find that the size and mass of the bulge can increase from 0.92 kpc to 1.9 kpc and from $4.7\times 10^{10}\, \rm M_\odot$ to $7\times 10^{10}\, \rm M_\odot$.

\end{abstract}

%{\color{purple} Notes: First draft.}




%%%%%%%%%%%%%%%%%%%%%%%%%%%%%%%%%%%%%%%%%%%%%%%%%%%%%%%%%%%%%%%%%%%%%%%%%%%%%%%%
%%%%%%%%%%%%%%%%%%%%%%%%%%%%%%%%%%%%%%%%%%%%%%%%%%%%%%%%%%%%%%%%%%%%%%%%%%%%%%%%
%(e.g., \citealp{}; \citealp{}; \citealp{})

%------------------ %%% Section 1 %%% --------------------%

\section{Introduction} \label{sec:intro}

%--------% The SMBHs and bulge %---------%

Supermassive black holes (SMBHs) have been discovered in most galaxies (e.g., \citealp{lynden1969galactic,kormendy1995inward,magorrian1998demography}), suggesting that the existence of SMBHs may be associated with their host galaxies. Growing evidence shows that the mass of an SMBH is correlated to its galactic bulge, including luminosity (e.g., \citealp{kormendy1995inward,wu2001black, graham2007log, marconi2003relation}), stellar mass (e.g., \citealp{mclure2002black, haring2004black, graham2001correlation}), size \citep{ferrarese2002beyond, lauer2007masses}, and dynamics and velocity dispersion $\sigma$ (e.g., \citealp{ferrarese2000fundamental, tremaine2002slope, aller2007host,soker2011correlation, king2015powerful}). These observational results exhibit strong connections between the SMBHs and their host galaxies (e.g., \citealp{wandel2002black, feoli2011smbh, kormendy2013coevolution}).

%-------% Super-massive black hole growth problem %--------%

The SMBH of the Milky Way (MW), known as “Sagittarius A$^\ast$”, has a mass of $(4.1\pm 0.6) \times 10^{6}  \rm\, M_{\odot}$ \citep{ghez2008measuring}. The maximum accretion rate of the Sagittarius A$^\ast$ is $8 \times 10^{-5}\ \rm\, M_\odot\ yr^{-1}$, which has been determined from X-ray and infrared research \citep{quataert1999accretion}. However, \citet{mortlock2011luminous} and \citet{venemans2015identification} discovered SMBHs of $10^9 \rm\, M_\odot$ in the early universe at $z > 6.5$. To grow an SMBH with a mass of one thousand times larger than that of Sagittarius A$^\ast$, high accretion rates are required in a very short period (The universe's age is only $\simeq \rm 0.83\ Gyr$ at $z=6.5$). How an SMBH grows in a short time and how the host galaxy provides a suitable environment for the growth remain unknown.
One possible explanation is that an SMBH grows through galaxy mergers \citep{volonteri2005rapid}, which supply the SMBH with a new gas reservoir. In a fully developed {isolated} disk galaxy, the dark matter, {gas cloud, and stars have reached the state of relaxation; the galaxy is in quasi-dynamical equilibrium.} Therefore, most of the gas and stars move along {their stable orbits without migrating} toward the galactic center. {In contrast}, during a major merger between two galaxies, the tidal forces from their interaction can induce strong, non-axisymmetric perturbations that disrupt the equilibrium of both galaxies and alter the motion of stars and gas. As a result, some stars and gas may be shifted towards the galaxy's center, forming a rich reservoir of gas that can fuel the growth of an SMBH during the phase of a galaxy merger. For example, an SMBH can rapidly grow during a wet merger \citep{hopkins2006unified} through a super-Eddington accretion \citep{takeo2019super}. However, how to maintain an enduring super-Eddington accretion remains unclear. Simultaneously, the feedback of SMBHs also self-regulates its accretion and star formation rates inside the galactic bulge \citep{springel2005black}. Simulations of galaxy merger by \citet{di2005energy} showed that an SMBH could grow from $4\times 10^{5} \rm\, M_\odot$ to $ 10^{8} \rm\, M_\odot$ in 1.5 Gyr, but growing an SMBH to $\sim 10^9 \rm\, M_\odot$ within 1 Gyr is still challenging.\par 

If the gas accretion powers the rapid growth of a galactic SMBH during major mergers of their host galaxies, such a substantial influx of gas should also drive a violent star formation activity at the galactic center. This so-called nucleus starburst of a star formation rate (SFR) $\sim 100 - 1000$ higher than normal galaxies have been observed in interacting galaxies (e.g., \citealp{joseph1985recent}, \citealp{schweizer2005merger}). Since stars form from molecular gas, we suspect that GMCs can also be used to grow an SMBH. GMCs contain dense molecular gas and have a more compact structure than diffuse gas. The difference in cloud masses between atomic gas clouds and GMCs leads to distinct dynamical friction effects from the surrounding medium in their host galaxy. Compact and massive GMCs are subject to stronger dynamical friction force from the background medium, consisting of gas, stars, and dark matter. This mechanism causes GMCs to migrate efficiently toward the galactic center during galaxy mergers. Therefore, we propose that GMC accretion can be a viable channel to facilitate rapid SMBH growth and generate a nucleus starburst in major mergers.

Due to the strong impact of a galaxy-galaxy collision, the original spiral structures of galaxies can be disrupted during major mergers. On the other hand, minor mergers are more promising for growing a galaxy's size \citep{bedorf2013effect} and contributing to a bulge's mass \citep{hopkins2010mergers}, producing irregular galaxies \citep{bournaud2005galaxy}. Furthermore, a minor merger may account for the inner components (inner discs and rings) in unbarred spiral galaxies \citep{eliche2011minor}. 

Therefore, we examine the possibility of SMBHs' rapid growth via accreting molecular gas in major and minor galaxy mergers using high-resolution galaxy simulations with the \texttt{GIZMO} code. The structure of this paper is as follows. Sec. \ref{section2} describes the simulation setup and the relevant physics required to model galaxy mergers. Then, the growth of SMBHs and the associate SF activities during galaxy mergers are presented in Sec. \ref{section3}. Finally, we discuss and compare the results with previous studies in Sec. \ref{section4} and conclude in Sec. \ref{section5}.

%%%%%%%%%%%%%%%%%%%%%%%%%%%%%%%%%%%%%%%%%%%%%%%%%%%%%%%%%%%%%%%%%%%%%%%%%%%%%%%%
%%%%%%%%%%%%%%%%%%%%%%%%%%%%%%%%%%%%%%%%%%%%%%%%%%%%%%%%%%%%%%%%%%%%%%%%%%%%%%%%

%---------------- %%% Section 2.1 %%% ---------------------%

\section{Numerical Methods}\label{section2}

\subsection{The \texttt{GIZMO} code}\label{sub2.1}

We perform our galaxy simulations using the mesh-free hydrodynamic code: \href{http://www.tapir.caltech.edu/~phopkins/Site/GIZMO_files/gizmo_documentation.html}{\texttt{GIZMO}}  \citep{hopkins2015new}. It is an open-source code that is based on the widely used cosmological code, \texttt{GADGET-2} \citep{springel2005cosmological}, for the domain decomposition and N-body algorithms. We employ the Meshless Finite-Mass (MFM) and Meshless Finite-Volume (MFV) methods in our {\texttt{GIZMO}} simulations. In the MFM and MFV methods, each particle is treated as a mesh-generating point that defines the volume. The physical quantities of each particle determine the fluid properties. The code solves the hydrodynamics equations by integrating the domain of each particle. Compared to other public codes in astrophysical simulations, \texttt{GIZMO} \citep{hopkins2015new} exhibits superior performance in conserving angular momentum, which is critical for modeling the dynamics of galaxy mergers. 

% Table of galaxy paramters.

\begin{deluxetable*}{ccccc}[htp]
	\tablecaption{Table of the initial conditions.}
	\tablehead{
		\colhead{Component} &
		\colhead{Physical Parameters} &
		\colhead{$\rm G_{rich}$} &
		\colhead{$\rm G_{poor}$} &
		\colhead{$\rm G_{dwarf}$}
		\\
	}
	\startdata
	Whole Galaxy & $M_{200}$\ [$\rm 10^{10} M_\odot$]   & 146 & 146 & 10\\
        		 & $V_{200}$\ [$\rm km \cdot\ s^{-1}$]  &  166.1& 166.1 & -- \\
            	 & Halo spin parameter ($\lambda$)  & 0.04& 0.04 &0.04\\
	\hline
	     	Gas  & Mass of atomic gas\ [$\rm{M_\odot}$] & $6.2\times 10^{10}$ & $1.5\times 10^{10}$ & $1.5\times 10^{9}$     \\ 
	     		& Mass of molecular gas\  [$\rm{M_\odot}$]   & $8\times 10^{10}$ & $6\times 10^9$ & $2\times 10^{8}$  \\ 
	     	    & radial cut\ [kpc]   &  15& 15&10 \\ 
	     	    & Mass of an atomic particle  [$\rm{M_\odot}$]  & $2\times 10^4 $& $5\times 10^3$ & $5\times 10^3$\\
 			    & Mass of a molecular particle  [$\rm{M_\odot}$] &$ 10^6$ &$ 10^6$&$ 10^6$ \\
 			    & Distribution forms & &{Exponential disk+sech-z}\\
                & Mean metallicity of the particles [$\rm Z_\odot$] & & $2\times 10^{-6}$\\
                & Gravitational softening lengths [kpc]& & 0.001\\
 	\hline
    Dark Matter & Mass\  [$\rm{M_\odot}$]    &  $1.2\times 10^{12}$ & $1.2\times 10^{12}$& $9.5\times 10^{10}$    \\ 
 			    & radial cut\ [kpc]  & 300& 300&100   \\     
 			    & Concentration parameter& 13& 13&4.4 \\
 			    & Mass of a DM particle [$\rm{M_\odot}$] &$5 \times 10^6$& $5 \times 10^6$ &$5 \times 10^6$\\
	            &  Distribution &   & NFW profile &  \\
                & Gravitational softening lengths [kpc]& & 0.05\\
	\hline
	Stellar Disk & Mass\ [$\rm{M_\odot}$]   & $1.1\times 10^{11}$ & $2.3\times    10^{11}$&$2.3\times 10^{9}$   \\ 
                         & radial cut\ [kpc]  & 20& 20 &10   \\ 
 			             & Mass of a star particle [$\rm{M_\odot}$] & $10^6$& $10^6$&$10^6$ \\
                &    Distribution &    & Exponential disk+exponential-z &  \\
                & Mean metallicity of the particles [$\rm Z_\odot$] & & 0.01\\
                & Gravitational softening lengths [kpc]& & 0.02\\
    \hline 
    Stellar Bulge & Mass\ [$\rm{M_\odot}$]  &  $8\times 10^{9} $ & $1.7\times 10^{10} $& $1\times 10^{9}$  \\ 
                          & radial cut\ [kpc]   &   2& 2&0.5\\ 
 			              & Mass of a star particle [$\rm{M_\odot}$] &  $10^6$& $10^6$&$10^6$\\
                          &  Distribution    &   & Einasto profile &   \\
                          & Mean metallicity of the particles [$\rm Z_\odot$] & & 0.001\\
                          & Gravitational softening lengths [kpc]& & 0.02\\
	\enddata
	\tablecomments{The initial mass of an SMBH seed in each galaxy model is not an input parameter. Instead, this seed mass is simulated based on the recipe discussed in Sec. 2.4.}
	\label{tbl:iso1}
\end{deluxetable*}

%---------------- %%% Section 2.2 %%% ---------------------%

\subsection{Simulation setup}\label{sub2.2}

We create pre-merging galaxies using the MW as a reference and divide them into gas-poor and gas-rich galaxies. The gas-poor galaxies, with a mass of $\sim 1.46 \times 10^{12} \rm\, M_\odot$, are modeled based on parameters obtain from Gaia observations \citep{brown2016gaia,li2016modellin,posti2019mass} and the N-body study by \citet{fujii2019modelling}. On the other hand, the gas-rich galaxies adopt the same parameters as the MW but with an artificially increased gas mass of $1.42\times 10^{11} \rm\, M_\odot$. The population of gas-rich galaxies likely dominates among high-redshift galaxies of $z> 2$ \citep{solomon2005molecular}. 

Herein, both the gas-poor and gas-rich galaxies here are called {$10^{12} \rm\, M_\odot$ disk} galaxies. In addition to {these} galaxies, we generate dwarf galaxies with mass $\sim 1 \times 10^{11} \rm\, M_\odot$ for minor mergers. The detailed parameters of these three types of galaxies are listed in Table \ref{tbl:iso1}. Then, we use the \texttt{DICE} code \citep{perret2014evolution} with input parameters from Table \ref{tbl:iso1} and create the initial galaxies for evolution in the \texttt{GIZMO} code. The simulation considers the following scenarios:
 \begin{enumerate}
     \item Model $\bf \rm  M_{poor}$ denotes a major merger of two gas-poor galaxies.
     \item Model $\bf \rm  M_{rich}$ denotes a major merger of two gas-rich galaxies.
     \item Model $\rm  M_{dwarf}$ denotes a minor merger of one gas-poor galaxy with two dwarf galaxies.
\end{enumerate}

For the major mergers, we place two galaxies in a $(100\ \rm{kpc})^3$ cubic simulation box, and their separation is 30\ kpc with an approaching velocity of 180 $\rm km\ s^{-1}$. The two galaxies will collide in 0.1 billion years after the simulation begins. For the minor merger case, we place a gas-poor galaxy and two dwarf galaxies in the simulation box. The gas-poor galaxy is located at the center of the box, while two dwarf galaxies are separately placed 37 kpc and 55 kpc away from it. The two dwarf galaxies approach the gas-poor galaxy with a velocity of 120 km s$^{-1}$. The initial conditions for these three scenarios are illustrated in Table \ref{tbl:merger}. 
 
We plan to initially evolve three models for five billion years (5 Gyr). As the simulations evolve, their time step shrinks rapidly at the late time due to the complicated interplay between the gas flow and SMBH/SNe feedback. It makes the simulation very computationally expensive and forces us to stop evolving the major merger simulations at $t \sim 2$ Gyr and $t \sim 3.5$ Gyr for the minor merger case. The simulation results are analyzed with python packages \texttt{SciPy} \citep{virtanen2020scipy}, \texttt{pylab} \citep{hunter2007matplotlib}, and \texttt{h5py} \citep{collette2013python} with data visualization using \texttt{Matplotlib} \citep{hunter2007matplotlib} and the \texttt{SPLASH} code \citep{price2007splash}.\par

% Table of galaxy merger parameters.

\begin{deluxetable}{ccccc}[htp]
	\tablecaption{Table of the galaxy merger parameters.}
	\tablehead{
		\colhead{Name} &
		\colhead{$d_{\rm ini}\ [\rm kpc$]} &
		\colhead{$v_{i}\ [\rm km\cdot s^{-1}]$} &
		\colhead{$s\ [\rm deg]$} &
		\colhead{$i\ [\rm deg]$}
		\\
	}
	\startdata
	 $\rm M_{poor},\quad \rm M_{rich}$  & 30  & 180 & 0 & -90\\
	\hline
	 $\rm M_{dwarf}$  & 37 & $120$ & $45$ & $45$     \\ 
	     	                      & 55  & $120$ & $0$ & $90$  
	\enddata
	\tablecomments{$d_{\rm ini}$ is the initial separation, $v_i$ is the initial velocity, $s$ is the spin angle, and $i$ is the inclination angle.}
	\label{tbl:merger}
\end{deluxetable}

%---------------- %%% Section 2.3 %%% ---------------------%

\subsection{ISM Physics in \texttt{GIZMO} }\label{sub2.3}

In reality, the interstellar medium (ISM) structure is incredibly complex, with different phases of gas, dust, and other baryonic components. Modeling the cold and dense regions is challenging because of the extremely small timesteps required for hydrodynamic simulations \citep{marinacci2019simulating}. To overcome this challenge, a common solution is to use an “effective equation of state (eEOS)”\citep{springel2003cosmological,vogelsberger2013model} that treats the multi-phase ISM gas as an entity and calculates its EOS based on the contribution from each phase of the ISM. Therefore, we use a widely-used two-phase sub-grid model based on \citet{mckee1977theory,springel2003cosmological} added molecular gas particles to model the galactic ISM. This model includes the hot atomic gas ($T=3\times 10^{8}$ K) for the SN ejecta, cold atomic gas ($T=10^{3}$ K) for the diffuse ISM, and cool molecular gas ($T=30$ K) for the GMCs. The cooling and heating of the two-phase atomic gas are modeled through ionization and recombination of hydrogen/helium, photo-electrons, collisional excitations, bremsstrahlung, cosmic rays, and Compton scattering for temperatures above $10^4$ K \citep{hopkins2018fire, faucher2020cosmic}. For the metal cooling, \texttt{GIZMO} uses a cooling table of 11 elements (H, He, C, N, O, Ne, Mg, Si, S, Ca, and Fe) calculated by \texttt{CLOUDY} \citep{ferland1998cloudy,wiersma2009effect}. For molecular gas cooling, additional fine-structure and molecular cooling functions from \citet{robertson2008molecular} extend the cooling curves down to $\sim 10$ K \citep{hopkins2018fire}.  

The gas in \texttt{GIZMO} comprises atomic and molecular gas particles. We set the mass ratio between molecular and atomic gas, $M_{\rm H2}/M_{\rm HI} \sim 1.3$ for the gas-rich merger based on the semi-analytic models \citep{popping2014evolution}, and $ \sim 0.4$ for the gas-poor merger based on the observation of our Milky Way \citep{brown2016gaia,li2016modellin,posti2019mass}. Because the simulation did not consider the exchange processes between atomic and molecular gas, an atomic gas particle never turns into a molecular particle, and vice versa.\par
ISM Observations \citep{solomon1979giant, solomon1980giant} suggested that most galactic molecular gas resides in GMCs. For simplicity, we assume all molecular gas in the simulation belongs to massive GMC particles, and each particle mass is set to $\sim 10^6 \rm\, M_\odot$ based on GMC observations \citep{cohen1985molecular,
solomon1987mass}. Meanwhile, each atomic gas particle has a mass of $\sim 10^3-10^4 \rm\, M_\odot$, which is $\sim 100$ times less than a molecular gas particle. The mass difference between atomic and molecular gas particles can alter their kinetics through dynamical friction. This force arises when an object moves through the galaxy and experiences a drag force due to the gravitational pull from gas, stars, and dark matter in the background. In sum, the dynamical friction can differentiate the dynamics of atomic gas and GMCs during major mergers. Section \ref{sub4.1} will delve into how this process takes place. 
 The galactic star formation in the simulation follows the star formation model of \citet{springel2003cosmological}, which only allows stars to form from molecular gas particles. The gas particles with a density $>100 \rm\ cm^{-3}$ and satisfy other star formation criteria (e.g., the Jeans criterion); these gas particles are then converted into star particles with a star formation efficiency (SFE) based on the free-fall time of a gas particle divided by the SF timescale of 40 Myr used in the code. We use the stellar feedback recipe from the AGORA project \citep{kim2016agora}, which applies “pure thermal energy dump” feedback: mass, metals, and thermal energy injected locally in a simple kernel-weighted fashion around the star particles.
 
 Once the star particles form from molecular gas, they immediately produce strong stellar feedback. Besides the thermal feedback, the algorithm assumes the newly forming star particles to have an IMF-averaged number of core-collapse SNe, which all explode at once when the age of a star particle turns into 5 Myr old and assumes each core-collapse SN releases $10^{51}$ erg explosion energy and 14.8 $\rm M_\odot$ ejecta containing 2.6 $\rm M_\odot$ of metals with a solar abundance.

%---------------- %%% Section 2.4 %%% ---------------------%

\subsection{Modeling the physics of SMBHs} \label{sub2.4}

Similar to previous studies, we use a sub-grid prescription \citep{hopkins2011analytic} to address the physical processes in an SMBH. We permit BH seeds {\citep{hoyle1962nature,rees1984black,chen2014general,woods2019titans}} to be formed in the galaxy's central region and allow them to merge to form an SMBH seed which accretes mass using the sub-grid accretion model from \citet{hopkins2010massive}. {These BH seeds are thought to originate from the collapse of  supermassive stars of mass $\approx 10^4 - 10^6 \rm\, M_\odot$ formed in the atomic cooling halos of mass $\sim 10^9 \rm\, M_\odot$ at $z \sim 10-15$ \citep{begelman2010evolution,inayoshi2020assembly,das2021formation}. The low-metallicity and high-dense environment are required to form these supermassive stars. Therefore, the criteria of BH seed formation are similar to the massive star formation in  galaxy simulation. Based on \citet{lamberts2016and} and \citet{grudic2018feedback}, the formation probability of BH seeds can be expressed as}

\begin{equation}
\frac{dP_{\rm seed}}{dM_\ast}=\frac{1}{M_n} [1-\exp(-\Sigma)]\ \exp(\frac{-Z}{0.01 Z_\odot}),\label{Eq.BH_seed}
\end{equation}
where $M_\ast$ is the stellar mass,  $M_n$ is a normalization parameter, $Z$ is the gas metallicity, and $\Sigma$ is the surface density in units of 1\ g cm$^{-2}$.\par

%\md{Initial BH seeds are formed in simulations based on the stellar evolution calculations performed by \citet{lamberts2016and}. Efficient metal line cooling occurs at around $Z\sim 10^{-2}$, where the densest gas has a cooling time much shorter than the free-fall time, i.e., $t_{\rm cool}\ll t_{\rm ff}$, leading to direct collapse \citep{grudic2018feedback}. On the other hand, \citet{grudic2018feedback} found that the critical surface density for BH seed formation is $\sim$ 0.1-1 g cm$^{-2}$, which is determined by the balance between the strength of stellar feedback and gravity.} Based \md{on the above recipe}, the formation probability of these seeds can be expressed;

Equation (\ref{Eq.BH_seed}) can be physically interpreted as indicating that BH seed particles tend to form in regions characterized by high surface density $\gg$ 1 g cm$^{-2}$ and low metallicity $\ll$ 0.01 Z$\odot$. Under these conditions, there is a high likelihood that a given gas particle will become a BH seed with a mass of $\sim 10^{6}$\rm\, M$\odot$. As listed in Table \ref{tbl:iso1}, the initial gas metallicity in both $\rm G_{rich}$ and $\rm G_{poor}$ is $2\times 10^{-6}$ Z$_\odot$, which soon will be increased by SN metals from deaths of massive stars within a few million years after the galaxies start to evolve. Therefore, the initial BH seeds only form at the innermost region ($<$ 0.3 kpc) at the beginning of the simulations due to the high-density and low-metallicity environment. These BH seeds then move toward the galactic center and eventually merge into a more massive seed, which we refer to as the SMBH seed.

The actual size of the accretion disk around the SMBH is estimated to be $2 - 10$ pc. Unfortunately, our simulation can only achieve the highest resolution of $\sim 20$ pc, so we cannot resolve the structure of the accretion disk. If a gas or star particle is located within 20 pc around the SMBH, then the code check criteria: 
(1). Is the particle gravitationally bound to the SMBH? 
(2). Is the apo-centric radius of the particle about the SMBH also $< 20$ pc? 
If both are true, the particle is immediately “captured” {by the SMBH}.

The physical properties of the accretion disk around an SMBH can be understood using the standard $\alpha$ disk model from \citep{shakura1973black}. However, the detailed process of mass transfer from the disk to the SMBH can only be understood through GRMHD simulations, which is far beyond the scope of this study. Therefore, we assume a steady-state accretion disk and allow $\dot{M}_{\rm BH} = \dot{M}_{\rm disk}$.

An accreting SMBH grows in mass and posits strong feedback to its surroundings. The intrinsic bolometric luminosity is $L_{\rm{bol}} \equiv \epsilon \dot M_{\rm {BH}}c^2$, where $\epsilon \sim 0.1$ is the radiative efficiency. This energy injection is equally distributed on the thermal energy of gas particles swirling around the SMBH \citep{springel2005modelling}. The injected energy immediately increases the temperatures and pressures of the associated gas particles that drive a strong outflow and impact the entire galaxy. Besides this thermal energy injection, we also consider the mechanical feedback from an SMBH by including an accretion disk wind model \citep{hopkins2016stellar}. The time-dependent mass and momentum from the wind are calculated with a mass loading factor of $\beta \equiv \dot M_{\rm {wind}}/{\dot M_{\rm BH}}=0.5$ and wind velocity $v_{\rm wind}=30000\ \rm{km\ s^{-1}}$ based on observation \citep{moe2009quasar,borguet2012major}. Similar to thermal energy, the mass and momentum from the disk wind are injected into the gas particles swirling around the SMBH. 

\subsection{Defining the Bulge size}

A galactic bulge contains a group of old stars in the center of a spiral galaxy. The bulge properties are highly related to its central SMBH (e.g., $M-\sigma$ relation in \citet{ferrarese2000fundamental} and bulge mass relation in \citealp{haring2004black}).  The size and mass of a galactic bulge can be determined by fitting a profile of stellar density distribution with the sersic profile form \citep{sersic1968atlas}:

\begin{align}\label{Eq:bulge_fitting}
      I(R) & = I_e \exp\bigg \{-(2n-0.327)[(\frac{R}{R_e})^{\frac{1}{n}}-1]\bigg \}  \notag \\
              &  +I_{S} \exp (-\frac{R}{R_S}),
\end{align}
where $I(R)$: the intensity at radius $R$, $I_e$: the effective intensity of the bulge, $n$: the sersic index, $R_e$: the effective radius of the bulge, $I_S$: the effective intensity of the disk, $R_S$: the effective radius of the disk.

During the galaxy merger, the bulge structure is poorly defined. Therefore, we adopt two methods to fit the bulge; if the stellar dynamics of the bulge are in equilibrium, we fit the bulge by assuming an SMBH in the bulge center. During a minor merger, the bulge stars take longer to reach equilibrium than during major mergers. Thus, we select the brightest region as the bulge center for fitting.

%---------------- %%% Section 3 %%% ---------------------%
\newpage
\section{Results}\label{section3}

% Figure 1 black hole growth

\begin{figure*}[htp]
\centering
\plotone{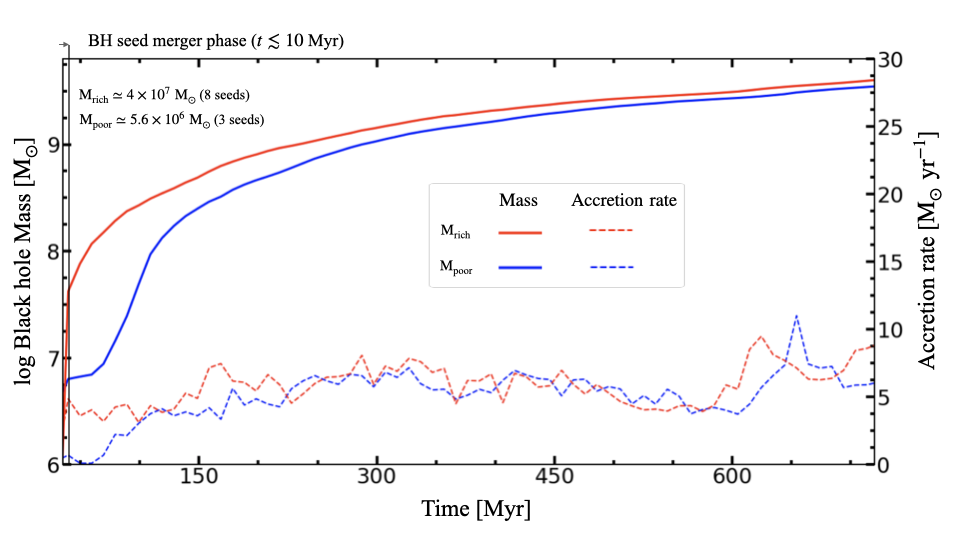}

\caption{The evolution of SMBHs in the $\rm M_{poor}$ and $\rm M_{rich}$ models. In the $\rm M_{poor}$ model, three BH seeds merge into an SMBH seed with a mass of $\rm 5.6\times 10^6\ M_\odot$, while in the $\rm M_{rich}$ model, eight BH seeds merge into an SMBH seed with a mass of $\rm 4\times 10^7\ M_\odot$. The average accretion rate is $5\ \rm M_\odot\ yr^{-1}$ for the $\rm M_{poor}$ model and $7\ \rm M_\odot\ yr^{-1}$ for the $\rm M_{rich}$ model. The SMBH feedback-accretion interaction leads to an oscillatory pattern in the accretion rates.}
\label{fig:Mbh growth}
\end{figure*}

% These SMBHs can grow rapidly with high accretion rates ($ > \rm 1\,  M_\odot\ yr^{-1}$), indicating that significant amounts of gas are falling into the central region to fuel the SMBHs. 
% Figure 2

\begin{figure*}[htp]
\plotone{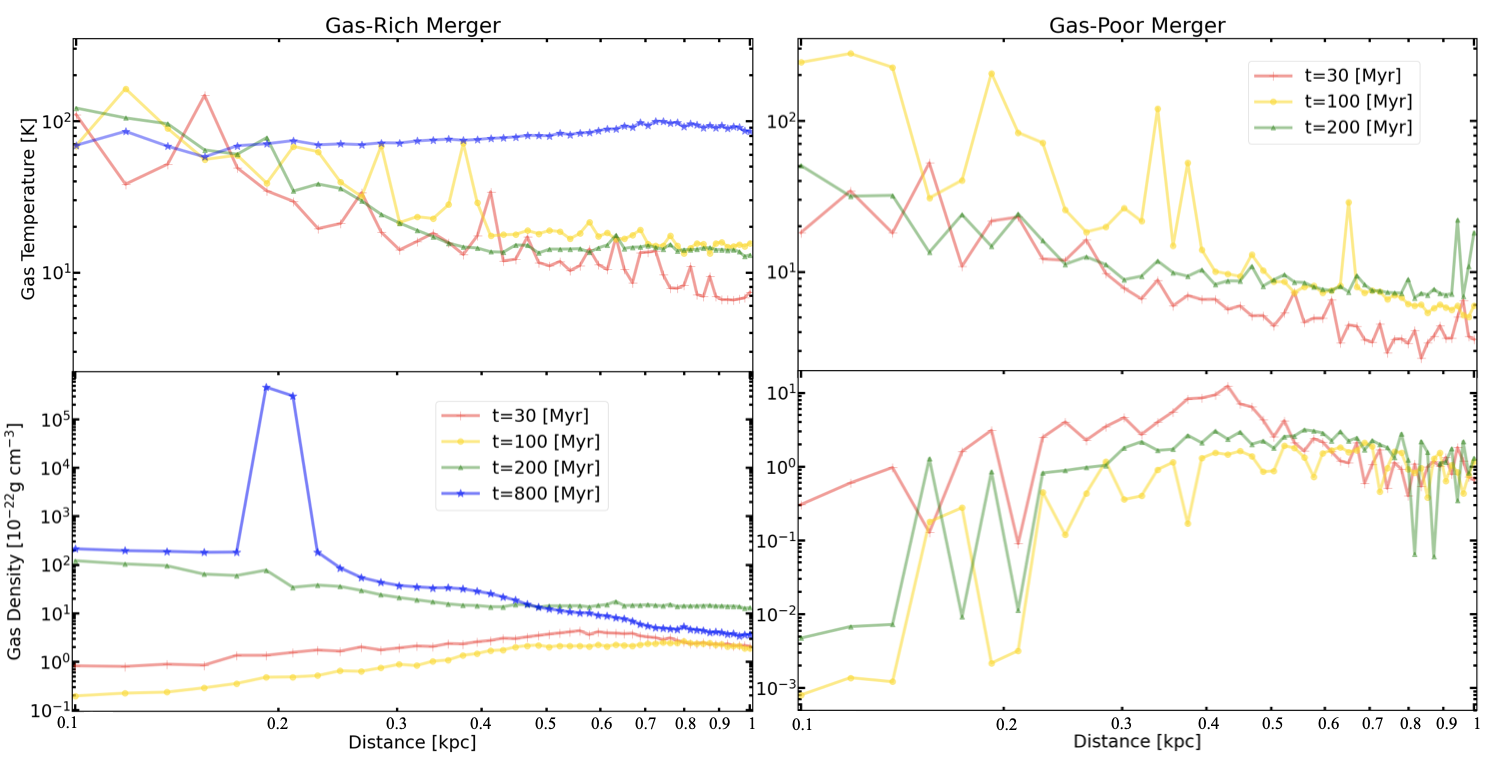}
\caption{1D projected gas temperature and density profiles around the SMBH. The red, yellow, green, and blue lines show the radial profiles at t=30, 100, 200, and 800 Myr, respectively. We calculate these mass-weighted profiles by projecting the 3D structure of density and temperature into a 1 kpc sphere centered at the SMBH.}
\label{fig:Mbh cond}
\end{figure*}

% Figure 3

\begin{figure*}[htp]
\plotone{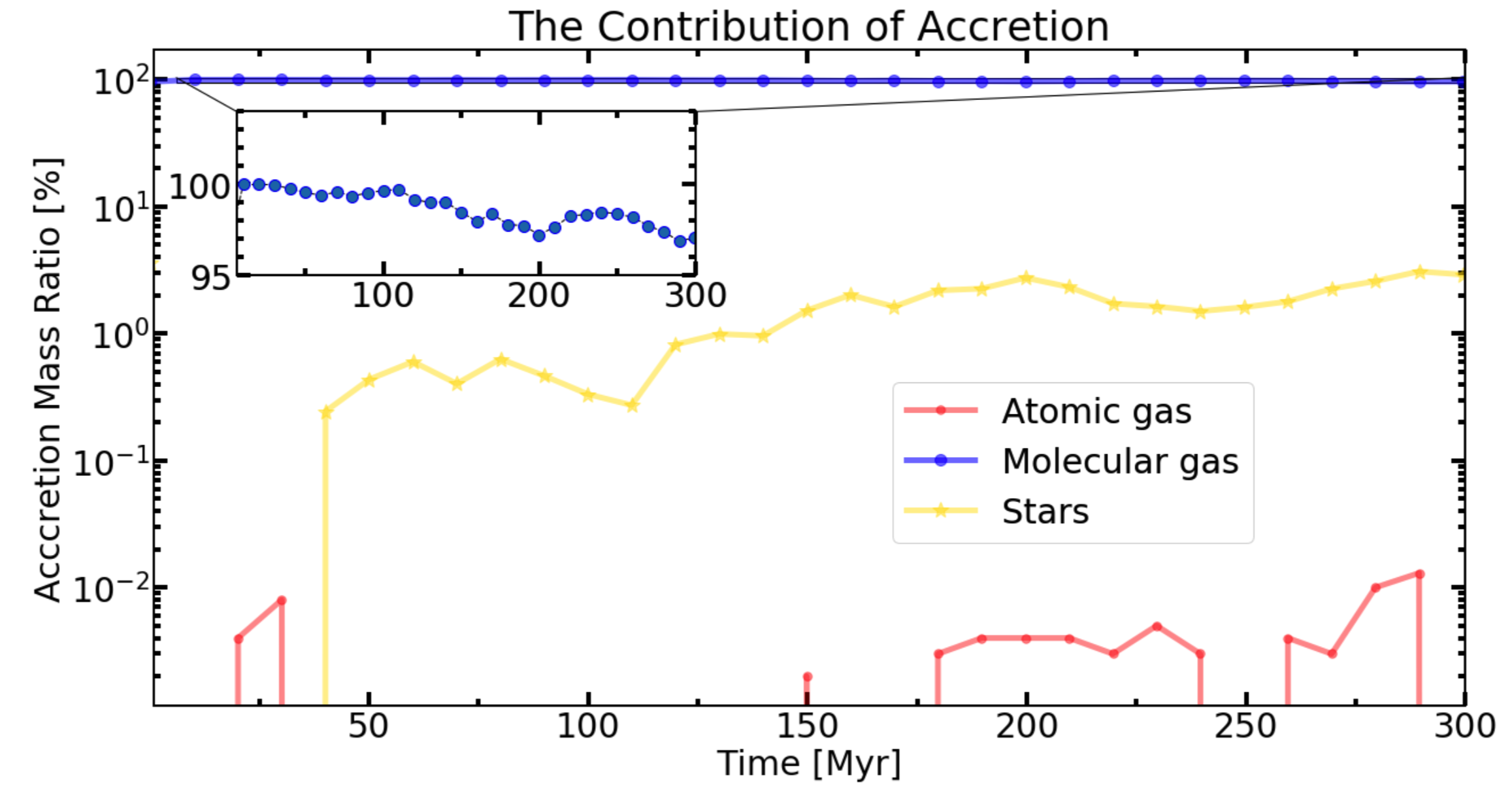}
\caption{The evolution of accreted compositions is shown by the color curves, representing the accretion of atomic gas, molecular gas, and stars onto the SMBH. It is evident that molecular gas contributes more than $95\%$ of the accretion mass, with the remaining $< 5\%$ coming mostly from stars. The contribution of atomic gas is minimal, accounting for less than $0.01\%$ of the SMBH accretion. Hence, the accretion of molecular gas from giant molecular clouds (GMCs) dominates the mass accretion onto the SMBH.}
\label{fig:contribution}
\end{figure*}

%---------------- %%% Section 3.1 %%% ---------------------%

\subsection{Rapid Growth of SMBH during the period of Major Mergers}\label{sub3.1}

First, we show the mass evolutionary tracks of SMBHs for the major merger of galaxies in the first Gyr in Figure \ref{fig:Mbh growth}. Overall, the average mass accretion rate is $\sim 5\ \rm\, M_\odot\ yr^{-1}$ and $\sim 7\ \rm\, M_\odot\ yr^{-1}$ for the $\rm M_{poor}$ and $\rm M_{rich}$ models, respectively. The mass accretion increased the mass of the SMBHs from $\sim 10^{7}\, \rm M_\odot$ to $10^{9}\, \rm M_\odot$ within 300 Myr. 
In the simulation, the Eddington limit is used to cap the maximum SMBH accretion rate, so the super-Eddington accretion cannot occur through the gas accretion. However, we find an extremely rapid growth of SMBH in the first 10 Myr in Figure \ref{fig:Mbh growth}. Although this accretion rate behaves like a super-Eddington, it is actually driven by the merger of initial BH seeds instead of gas accretion. As mentioned in Sec. \ref{sub2.4}, multiple BH seeds can form in the inner region of $\leq$ 0.3 kpc at the beginning of the simulation. We find eight BH seeds formed in the ${\rm M_{rich}}$ run and three BH seeds formed in the ${\rm M_{poor}}$ run. After the BH seeds form, they move towards the galactic center and finally merge to form the SMBH seed within 10 Myr. The mass of the SMBH seed in the ${\rm M_{poor}}$ model is $\rm 5.6\times 10^6\ M_\odot$ and in the ${\rm M_{rich}}$ model is $\rm 4\times 10^7\ M_\odot$.

The evolution of the mass accretion rate of the SMBHs in Figure \ref{fig:Mbh growth} shows oscillatory patterns which are likely driven by the SN or BH feedback. We estimate the energy injection rates from SNe and the SMBH to find the dominating feedback driving the oscillatory accretion rates. Based on Figure \ref{fig:Mbh growth}, the average accretion rate of SMBH is $\dot m \sim 5 \rm\ M_{\odot}\ $yr$^{-1}$, so the total thermal energy dumps to the SMBH surrounding area has the form $E_{\rm BH}\simeq L_{\rm bol} \simeq \epsilon \dot m c^2 \simeq 3 \times 10^{46}\ \rm erg\ s^{-1}$, assuming $\epsilon \simeq 0.1$. For the SNe feedback during starburst, we assume the typical explosion energy of an SN, $10^{51}$ erg. The energy injection is only $\sim 3 \times 10^{43}\ \rm erg\ s^{-1}$ for the SNe feedback by assuming the SFR in a starburst galaxy of $\sim 100\ \rm M_{\odot}\ $yr$^{-1}$ with the Chabrier IMF \citep{chabrier2003galactic} and a galactic SNe rate of $\sim 0.01 $\ yr$^{-1}$ \citep{rozwadowska2021rate}.
The SMBH feedback is $\sim 100 - 1000$ times stronger than the SN feedback. Therefore, the SMBH feedback is responsible for making the oscillatory patterns on the accretion rates of an SMBH.

The host galaxies' environment can affect the growth of the SMBH. For example, the SMBHs in gas-rich galaxy mergers can grow larger than those in gas-poor galaxy mergers. In Figure \ref{fig:contribution}, we show the compositions of accretion history. We find that the molecular gas contributes to $> 95\%$ of the accreted mass and plays a major role in fueling the rapid growth of an SMBH. Meanwhile, stars and atomic gas only contribute to $<5 \%$ of mass accreted. The above result demonstrates that the critical ingredient of the rapid growth of SMBHs in the simulation is the accretion of molecular gas. 

%The seperation between is getting smaller than 1 kpc.

% Figure 4

\begin{figure*}
\plotone{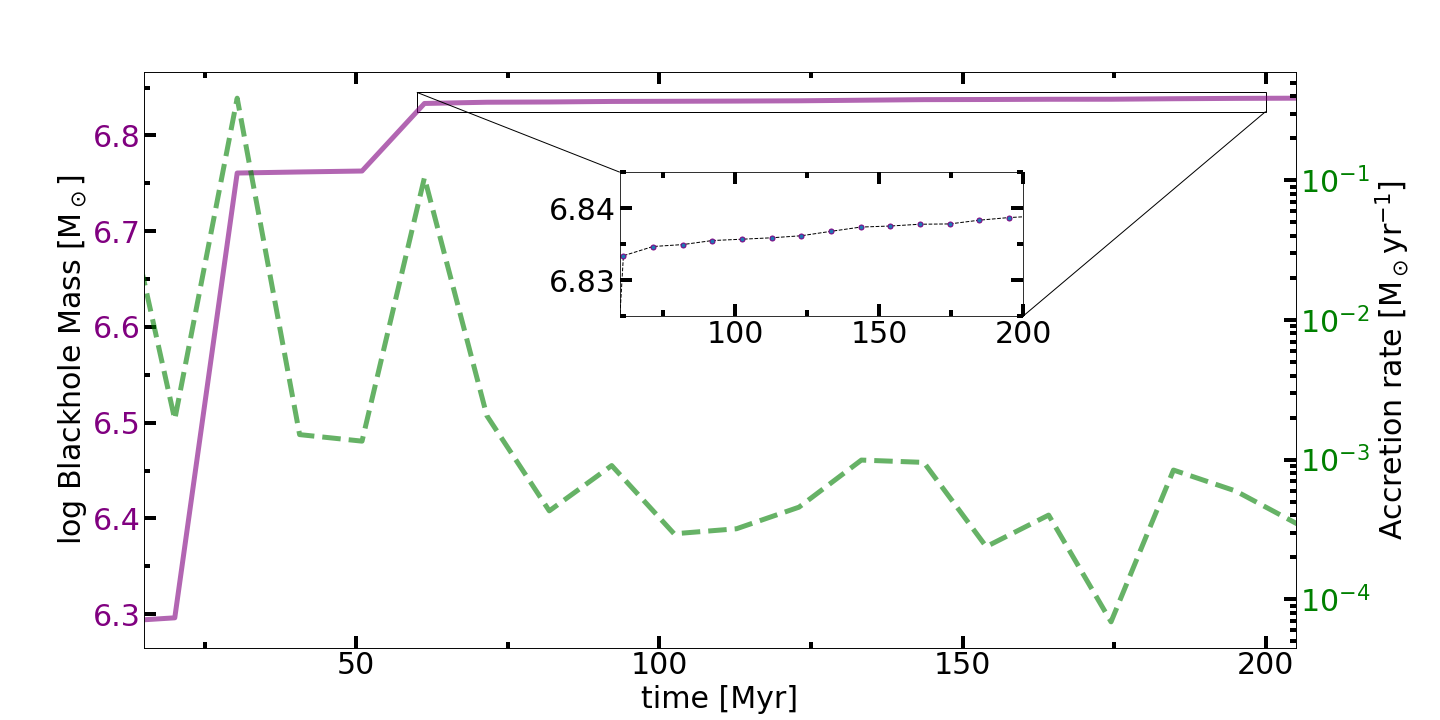}
\caption{The evolution of SMBH mass in the model with a particle mass of $10^4 \rm\, M_\odot$ for both molecular and atomic gas. The purple line indicates the SMBH mass evolution, while the green dashed line represents the SMBH mass accretion rate. The initial rapid growth of the BH seed from $2\times 10^{6} \rm\, M_\odot$ to $5.6\times 10^{6} \rm\, M_\odot$ in the first 10 Myr is due to the merger of BH seeds discussed in Section \ref{sub2.4}. However, because the molecular gas particle has the same mass as the atomic gas particle, the migration timescale for the “molecular gas particle” toward the galactic center becomes $\sim 5$ Gyr. Consequently, after 200 Myr, the SMBH only grows to $6.3 \times 10^{6} \rm\, M_\odot$, indicating that inefficient gas accretion fails to drive a rapid SMBH growth.}
\label{fig:same mass}
\end{figure*}
 
 In comparison, we also perform another simulation identical to the gas-rich run but set the mass for the molecular and atomic gas both equal to $10^4 \rm\, M_\odot$. As shown in Figure \ref{fig:same mass}, this equal-mass model fails to demonstrate a rapid growth of SMBH. The average growth rate in equal-mass run is $\sim 10^{-3} \rm\, M_\odot\ $yr$^{-1}$, which is consistent with the values found in \citet{di2005energy}.

%To check whether the dynamical friction between atomic and molecular particles can lead to the efficient growth of SMBH,

%---------------- %%% Section 3.2 %%% ---------------------%

 \subsection{Starbursts}\label{sub3.2}
\begin{figure*}
    \centering

    \plotone{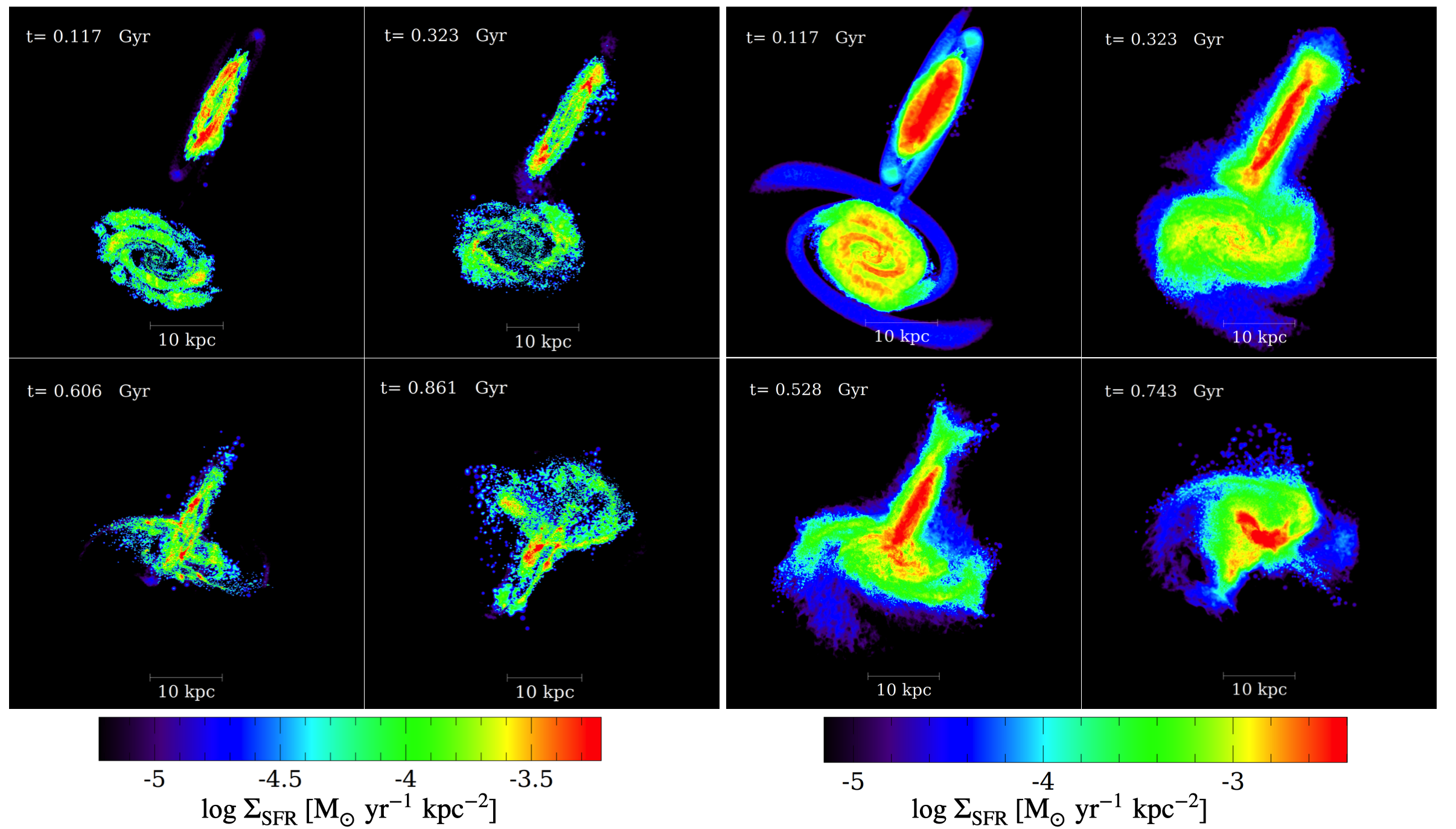}
    \caption{Left: The star formation rate density ($\Sigma_{\rm SFR}$) in the model $\rm M_{poor}$. The strong star formation regions shift from the spiral arms to the centers. Right: The star formation rate density ($\Sigma_{\rm SFR}$) in the model $\rm M_{rich}$. The molecular gas is more concentrated in the central region, and the SFR is enhanced, which drives the nucleus starburst.}
 \label{fig:SFD}
\end{figure*}

\begin{figure*}[htp]
\centering
\plotone{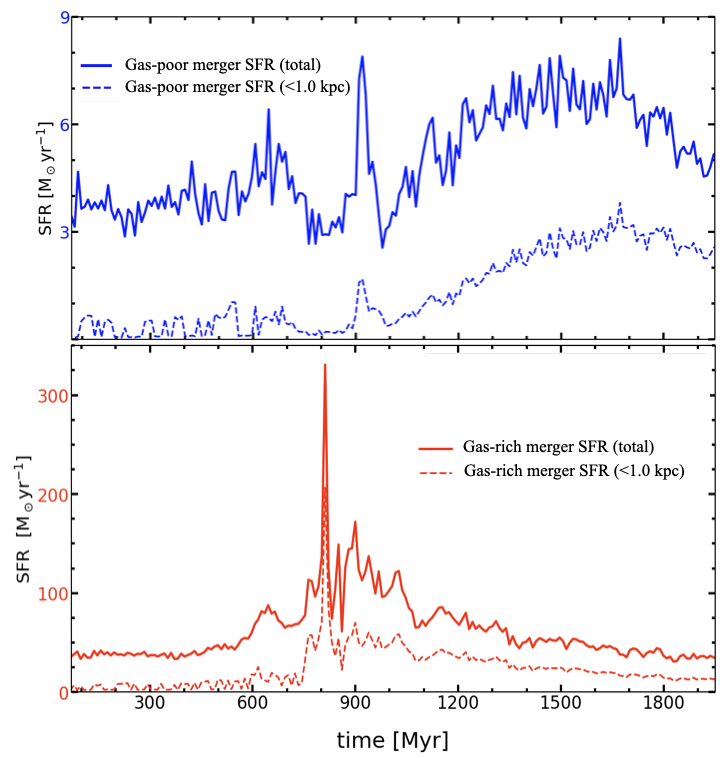}
  \caption{The SFR history during major mergers.  In $\rm M_{rich}$, the SFR reaches $\ge 100\rm  \, M_\odot\, yr^{-1}$ at $t \sim 800$ Myr and lasts for $\sim 150$ Myr.
   Such a violent SF event is known as the starburst phenomenon. The dashed lines represent the SFR in the central 1kpc region of the $\rm M_{poor}$ and $\rm M_{rich}$ models, respectively. The trend of the nucleus star formation is highly related to the starburst, implying that a large amount of molecular gas moves to the galactic center during this period.} 
  \label{fig:TSFRM}
\end{figure*}

At the beginning of our galaxy simulation, star formation mainly occurs in the spiral arms and the bulge that hosts GMCs until the two galaxies collide. The SFR distribution of the major merger cases is shown in Figure \ref{fig:SFD}. Many molecular-gas particles flow into the centers of two galaxies, triggering rapid star formation. In Figure \ref{fig:TSFRM}, a high star formation rate is observed at the central region of r < 1kpc, contributing to over 50\% of the star formation in the galaxy, and this high SFR lasts 150 Myr. In the simulation, we only allow star formation from molecular-gas particles; thus, the SFR distribution can reflect the importance of the dynamics of molecular gases. The starburst only occurs in the merger of gas-rich galaxies with a plentiful supply of molecular gas. 

%Star formation stems from molecular gas; therefore, the gas abundance affects the SFRs. In our result, the starburst only occurs in the merger of gas-rich galaxies.  

% Figure 5

% Figure 6

%representing the total SFRs of the $\rm M_{poor}$ and $\rm M_{rich}$ models, respectively.meanwhile the SFR in MW is $\rm 1\ M_\odot\, yr^{-1}$).
%---------------- %%% Section 3.3 %%% ---------------------%

\subsection{The bulge growth with minor merger}\label{sub3.3}

\begin{figure*}[htp]
    \centering
    \plotone{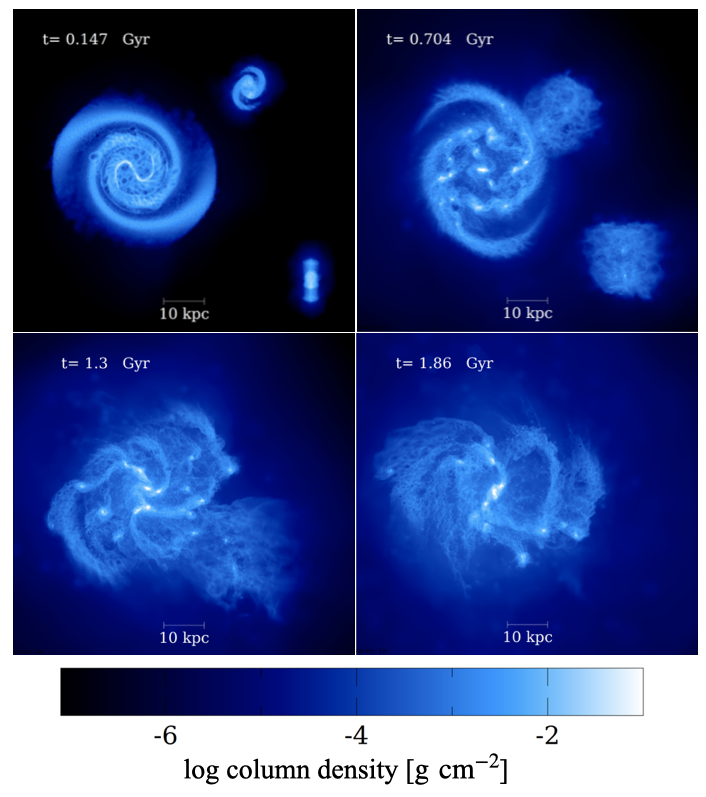}
    %\caption{(a) blah (b) blah (c) blah (d) blah}
  \caption{The evolution of minor mergers. The primary galaxy is a gas-poor galaxy, and we put two dwarf galaxies with different tilting angles to the galactic plane placed near the primary galaxy (see Table \ref{tbl:merger} for the angle information). Unlike the major mergers that disrupt the original spiral arms of the galaxy, minor mergers retain the original spiral arms of the primary galaxy.}
\label{fig:minor}
\end{figure*}

\begin{figure*}[htp] 
\centering
\plotone{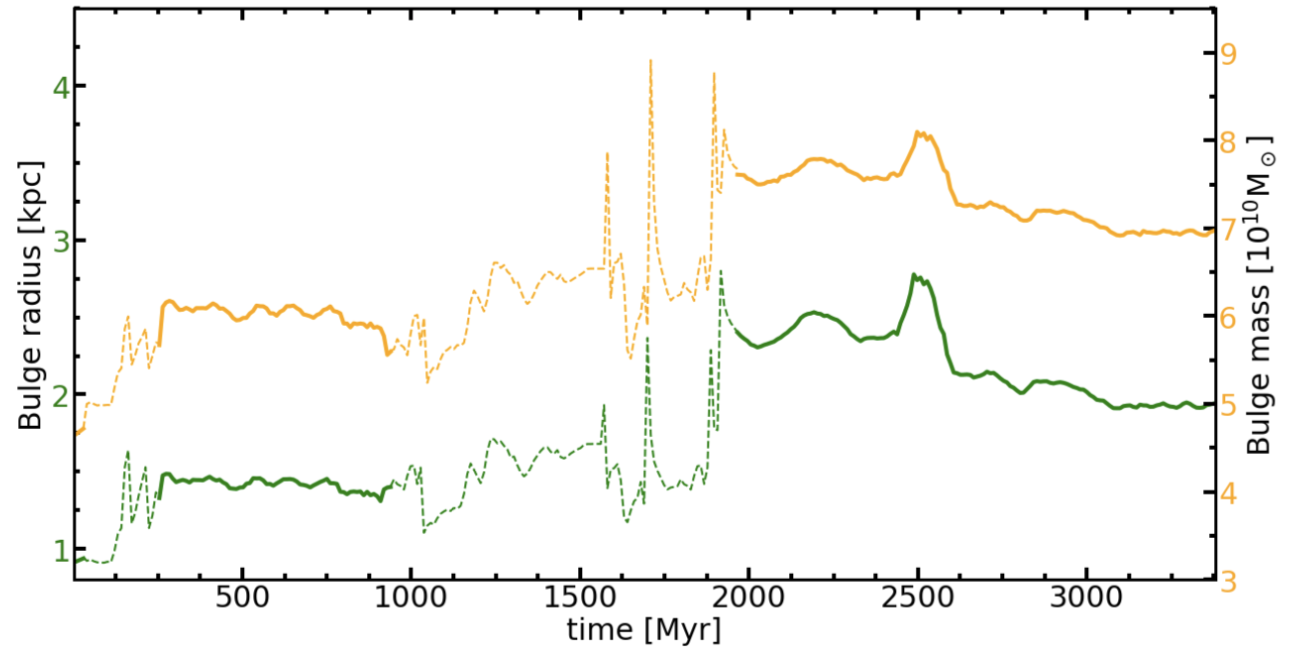}
 \caption{The bulge evolution in the $\rm M_{poor}$ model. The green line represents the bulge size. The initial size of the bulge in the $\rm M_{poor}$ model is 0.92 kpc, and it grows to 1.9 kpc after two minor mergers. The orange line denotes the total mass of the bulge. The initial mass of bulge in the $\rm M_{poor}$ model is $\sim 4.7 \times 10^{10}\ \rm\, M_\odot $, and it increases to $\sim 7.0 \times 10^{10}\ \rm\, M_\odot $ after minor mergers. The solid line sections from the bulge fitting refer to the SMBH as the bulge center. However, during minor mergers, the stellar density profile cannot fit well with the sersic function referring to the SMBH. Therefore, we select the position of highest $\Sigma_{\rm \ast}$ as the bulge center for fitting, and it yields the dashed line sections.}
\label{fig: bulge growth}
\end{figure*}

Since the bulge size for a detached galaxy is independent of the gas abundance, gas-poor and gas-rich galaxies have the same bugle size of 0.92 kpc. Moreover, the bulge sizes remain approximately constant during their evolution. A major merger cannot increase the bulge size of a spiral galaxy because this merger usually disrupts the original disk and bulge of galaxies. 
To investigate the growth of the bulge of a spiral galaxy, we consider minor mergers to avoid destroying the original structure of the primary galaxies. We consider the minor merger of a gas-poor galaxy with two dwarf galaxies with physical parameters from Table \ref{tbl:iso1} and run this model for 3.5 Gyr. Figure \ref{fig:minor} shows the evolution of bulge during the minor mergers. After two dwarf galaxies merge with the spiral galaxy, the bulge size increase by showing a larger bright central area in Figure \ref{fig:minor}. We then plot the 
evolution of bugle radius and mass in Figure \ref{fig: bulge growth} that shows the growths of bulge radius from 0.9 to $\sim$ 1.9 kpc and bulge mass from $4.7 \times 10^{10}\ \rm\, M_\odot$ to $7 \times 10^{10}\ \rm\, M_\odot$ at the end of the simulation. The average growth rate of the bulge mass is $\sim 30\ \rm M_\odot\ yr^{-1}$. The bulge of a disk galaxy can grow through minor mergers by swallowing the gas and dark matter of smaller galaxies. Therefore, newly-acquired dark matter will re-distribute onto the primary galaxy \citep{ferrarese2002beyond}. Thus, minor mergers can grow the bulge of the primary galaxy in mass and size without destroying the spiral structure.  

% Figure 8

%---------------- %%% Section 4 %%% ---------------------%

\section{Discussions}\label{section4}

%---------------- %%% Section 4.1 %%% ---------------------%

\subsection{Physics and caveat of the SMBH growth}\label{sub4.1}
Our results show that the SMBH in the {main} disk galaxy ($\sim 10^{12}\rm\, M_\odot$) can accrete GMCs to increase its mass via a major merger. This process allows the SMBHs' mass to increase from ${10^7} \rm\, M_\odot$ to $10^9\ \rm\, M_\odot$ within 300 Myr. Furthermore, the SMBHs in a gas-rich galaxy merger can attain larger masses than those in a gas-poor galaxy merger (Figure \ref{fig:Mbh growth}); this result well matches the results of \citet{di2005energy}, but our SMBHs' growth time is considerably shorter than that of \citet{di2005energy}.

We discuss the physics and uncertainties associated with our findings, indicating that an SMBH can experience rapid growth by accreting molecular gas. Based on \citep{chandrasekhar1943dynamical, carroll2017introduction, di2019black}, the dynamical friction in a galaxy can be expressed as: 
 \begin{align}\label{Eq:D_f}
     f \simeq 4\pi G^2 C\frac{M^2\rho}{v_M^2},
\end{align}
where $M$: the object's mass, $v_M$: the object's velocity, $\rho$: the density of the surrounding medium, and $C$ is a function that depends on how $v_M$ compares with the velocity dispersion of the surrounding medium. The dynamical friction is proportional to the mass squared of the object, as indicated by Equation (\ref{Eq:D_f}). Because a GMC particle is about 100 times heavier than an atomic gas in the simulation, the dynamical fiction for a GMC then becomes 10,000 stronger than atomic gas. Therefore, this strong force will drag the innermost molecular gas toward the galactic center and fuel the growth of an SMBH during major mergers.

\citet{jeffreson2018general} suggest the lifespan of a GMC of $\sim 10-100$ Myr that sets an upper limit in the migration time for a bulge GMC moving toward an SMBH. 
Based on Equation (\ref{Eq:D_f}), we estimate the migration timescale $t_m$ for a GMC moving from the stellar bulge to the galactic center; 
 \begin{align}\label{Eq:D_t}
\frac{d{v_M}}{dt_m} \simeq 4\pi G^2 C\frac{M\rho}{v_M^2} \Rightarrow t_m\simeq \frac{{v_M}^3}{12\pi G^2 C M\rho},
\end{align}
where $G$: the gravitational constant, $M$: the GMC's mass $\sim 10^{6} \rm\ M_{\odot}$, $v_M$: the GMC's drift velocity relative the galactic center, $\rho$: the matter density of the stellar bulge, and $C$ is $\sim 100$ for a typical stellar bulge. The matter density is $\rho=3M_b/4\pi R_b^3$ by assuming a spherical symmetry bulge of mass $ M_b \sim 10^{10} \rm\ M_{\odot}$ with a radius of $R_b \sim 1$ kpc.  The drift velocity of a GMC toward the galactic center takes its initial keplerian velocity $v_M \sim \sqrt{{GM_b}/{R_b}}$. Equation (\ref{Eq:D_t}) can be rewritten and yields $t_m$:
 \begin{align}\label{DF_t}
t_m\simeq \frac{1}{9C}\frac{M_b}{M}\sqrt{\frac{R_b^3}{GM_b}}\simeq 5\times 10^7 \rm yr \sim 50 \rm\ Myr, 
\end{align}
which is within the maximum lifespan of a GMC of $\sim 100$ Myr.
Therefore, GMCs can possibly arrive at the galactic center and feed an SMBH before they are dissipated. Some short-lived GMCs of lifespan $< 50$ Myr may dissipate before accreting onto the SMBH. However, these dissipated clouds may be reformed to GMCs around the galactic center ($R\sim 0.2$ kpc) \citep{morris1996galactic,genzel2010galactic,kauffmann2017galactic}. Furthermore, the successful rapid growth of an SMBH only requires accreting a fraction of halo gas rather than the entire gas from the host galaxy; our assumption of GMCs remains effective in demonstrating the phenomenon that GMCs migrate to the galactic center during galaxy mergers. In comparison, the migration timescale of the atomic-gas particle ($\sim 10^{4} \rm\ M_{\odot}$) is 5000 Myr, which hinders the accretion of atomic gas.

We argue that the massive GMCs have strong dynamical friction for efficient accretion onto an SMBH. However, each star particle has a mass of $\sim 10^{6} \rm\, M_\odot$, and each dark matter particle has a mass of $\sim 5 \times 10^{6} \rm\, M_\odot$; both also feel strong dynamical friction to drag them into the galactic center. Do star and dark matter particles contribute to the rapid growth of SMBH, too? The star particles have strong stellar feedback, which heats their surrounding environments and alters the mass distribution of gas. As a result, star particles' trajectories can be disturbed, which prevents these star particles from being swallowed by the SMBH. As for the dark matter accretion, the nature of dark matter is still unknown; whether their particle-like or wave-like behaviors can significantly affect their accretion physics, which is beyond the scope of this study. Therefore, we prohibit the accretion of dark matter onto the SMBHs in the simulation.\par

%---------------- %%% Section 4.2 %%% ---------------------%

\subsection{Possible destructive and reformation mechanisms of GMCs}

Besides the lifetime, several physical processes can destroy GMCs during their journey toward an SMBH. These processes include ram-pressure stripping, tidal disruption, and irradiation. In this section, we will discuss the effects of these mechanisms. To assess the impact of ram-pressure stripping on GMCs, we compare the magnitude of the ram-pressure acting on a GMC ($P$) to the gravitational binding energy density ($U_g$). The ram pressure is given by
\begin{equation}\label{Eq:ram}
P=\frac{1}{2}\rho v^2 \sim 1.4\times10^{-16} \rm\ kg\cdot m^{-1} \cdot s^{-2}, 
\end{equation}

where the ISM density, $\rho \sim 1\ {\rm particle}\ {\rm cm}^{-3}= 1.7\times 10^{-21}$ $\rm kg \cdot m^{-3}$, and the drift velocity of the molecular cloud toward the galactic center is $v \sim 20/10^5\ \rm pc\cdot yr^{-1}= 400\ m\cdot s^{-1}$ based on \citet{hsieh2017molecular}. In contrast, the gravitational binding energy density of a GMC  
\begin{equation}\label{Eq:grav}
U_g \approx (\frac{3GM^2}{5R})\frac{3}{4\pi R^3}\sim 4.3\times10^{-13}\ \rm kg\cdot m^{-1} \cdot s^{-2}, 
\end{equation}
where $G$ is the gravitational constant, $M$ is the GMC mass of $\sim 10^6 \rm\, M_\odot=2\times 10^{36}\ kg$ and $R$ is the GMC radius of $\sim \rm 100\ pc=3.1\times10^{18}\ m$. Because the binding energy of the GMC is about 1000 larger than the ram-pressure from the surrounding medium, the ram-pressure stripping effect of GMCs is neglectable. Irradiation effects are also minimal because dust and $\rm H_2$ in GMCs can help to shield from the external UV radiation field. Even if the mechanisms mentioned above disrupt some GMCs, they can be reformed through the compression of diffuse atomic gas in high-pressure environments created by the large-scale shocks from spiral arm passages or stellar feedback from massive stars and supernovae \citep{clark2005star,lee2017effect}. For tidal disruption, a portion of the GMCs debris can be pulled from a distance of 20 pc towards a 2 pc circumnuclear disk (CND) within $\sim 10^{5}$ years based on the observation of Sagittarius A$^\ast$ \citep{hsieh2017molecular}. In Seyfert galaxies, star formation can occur in CNDs that supports dense molecular gas to persist in the vicinity of the SMBH, and this CND-scale molecular gas can fuel AGNs \citep{izumi2016circumnuclear}.

%---------------- %%% Section 4.3 %%% ---------------------%

\subsection{Relations among SMBH mass, stellar mass, and bulge properties}

\citet{reines2015relations} found an empirical relation of SMBH mass and stellar mass in {S/S0 galaxies with classical bulges and elliptical galaxies},
\begin{align}\label{Eq:BHtoStar}
\log( \frac{M_{\rm SMBH}}{\rm M_\odot})=& (8.95\pm 0.09)
\notag \\
    &+(1.40\pm 0.21)\log(\frac{ M_{\ast}}{\rm 10^{11}\ M_\odot}).    
\end{align}

In the model $ \rm M_{rich}$, the SMBH has a mass of $\sim \rm 10^{10}\ M_\odot $ at $t \sim 2000$ Myr. Based on Equation (\ref{Eq:BHtoStar}), the corresponding stellar mass is $\sim  (6\pm 4) \times 10^{11}\rm\, M_\odot$.
Meanwhile, we obtain the total stellar mass of $1.5\times 10^{11}\ \rm M_\odot$ by summing up the mass of the existing old stars in the disk and bulge, as well as the mass of newly formed stars from Figure \ref{fig:TSFRM}.  The total stellar mass in $ \rm M_{rich}$  falls in the lower end at this empirical relation {, yielding an over-massive SMBH compared to the bulge mass.} At $t \sim 2000$ Myr, the rapid growth of SMBH is close to the end. However, the violent SF activities are still happening, as shown by the non-decaying SFR in Figure \ref{fig:TSFRM}. Missing the contribution of unfinished SF drop the total stellar mass to the lower end. Besides, the stellar mass of a galaxy can also increase through minor mergers (see Fig. \ref{fig: bulge growth}). Because events of gas-rich major mergers are rare, minor and/or gas-poor mergers can replenish the missing stellar mass. In the model $ \rm M_{poor}$, the mass of SMBH is $ \sim \rm 3\times 10^{9}\ M_\odot$ with a total stellar mass of $2.6\times 10^{11}\ \rm M_\odot$ at $t= 2000$ Myr. If we put the $ \rm 3\times 10^{9}\ M_\odot$ SMBH in Equation (\ref{Eq:BHtoStar}), the resulting stellar mass would be $ (1.8 \pm 1.2) \times 10^{11}\rm\, M_\odot$, which matches well with the result from the $ \rm M_{poor}$ model.  
In the simulation, we observed two kinds of active star formation. During the merger of gas-poor galaxies, their SFR increases by a factor of 2 to 3, which agrees well with the observational results of \citet{pearson2019effect} for low-redshift galaxies. However, during the merge of the gas-rich galaxies, their SFR significantly increases and exhibits a nucleus starburst shown in Figure \ref{fig:TSFRM}. The timescale for the starburst is 150 Myr, which is close to the observational merger samples of $10^7$ yr to a few $10^8$ yr in \citet{cortijo2017spatially}. The large SFR stems from the merger-driven central star formation \citep{barnes1991fueling, mihos1995gasdynamics, saitoh2009toward, sparre2016zooming} when the fresh molecular gas flows into the galactic center and forms numerous stars. In comparison, if the original molecular model in \texttt{GIZMO} is used, no starburst is found during both gas-poor and gas-rich mergers. 
Our results also show that the bulge in a spiral galaxy can be grown through minor mergers \citep{bedorf2013effect}. For the minor merge of 
a gas-poor galaxy, its bulge radius grows from 0.9 kpc to 1.9 kpc, and mass increase from $4.7 \times 10^{10} \rm\, M_\odot$ to $7.0 \times 10^{10}\ \rm\, M_\odot$ after two minor mergers with dwarf galaxies. 
Because the Sagittarius A* is only $\sim 10^6 \rm\, M_\odot $, we suspect that our Milky Way may not have undergone any major mergers in its lifetime.

%---------------- %%% Section 4.4 %%% ---------------------%

\subsection{Limitations of models}\label{4.4}

Although our molecular gas model can realize the rapid growth of SMBHs, it still has several drawbacks to be improved. 
The accretion rate depicted in Figure \ref{fig:Mbh growth} represents an idealized maximum value based on the assumption of massive GMCs with a mass of $10^{6}\rm \, M_\odot$, rather than a more realistic mass distribution ranging from $10^{5}\rm \, M_\odot - 10^{6}\rm \, M_\odot$ \citep{rosolowsky2005mass,blitz2006giant}. Additionally, the assumption of a 100\% survival rate for GMCs during migration to the galactic centers is unrealistic, as GMCs can be subject to destructive mechanisms such as tidal disruption, ram-pressure stripping, and irradiation. However, replenishing dissipated GMCs through new molecular cloud formation is also possible. Furthermore, major mergers are rare events. Observations suggest that the number of major mergers is lower than that of active galactic nuclei (AGNs) in the redshift range of $0<z<3$ \citep{treister2010major}. Therefore, rapid SMBH growth via major mergers is more likely to occur in the high-$z$ universe, where interactions among galaxies are frequent.

Our simulation cannot model the formation and destruction of molecular gas from the first principles. As the first attempt, we adopt constant ratios of the molecular and atomic/ionized gas in the simulation. The bottleneck is to model the chemistry of multi-phase and multi-component ISM, and simultaneously differentiate the dynamics of molecular cloud and diffuse gas that remains beyond the envelope of our current ISM modeling. Recently, good progress has been made to improve the gas chemistry for galaxy simulations; for example, \citet{popping2014evolution} developed a molecular hydrogen formation recipes, and \citet{hopkins2018fire,hopkins2023fire} estimated the molecular hydrogen fraction based on the thermo-chemical equilibrium. 

% Figure 9.

\begin{figure*}[htp]
\centering
\includegraphics[width=\textwidth]{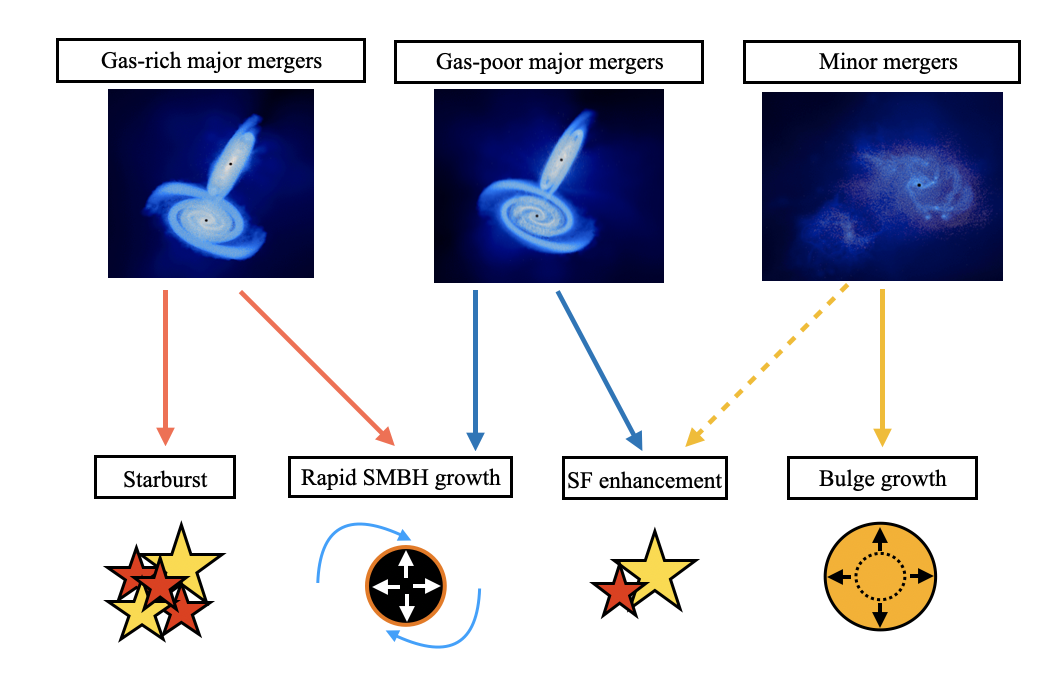}
\caption{This schematic illustrates the consequences of galaxy mergers. The red line represents the major merger of a gas-rich disk galaxy, the blue line represents the major merger of a gas-poor galaxy, and the yellow line represents the minor merger of a gas-poor galaxy. The solid lines represent the outcomes of the simulation, while the dashed lines depict possible evolution tracks. Following the rapid growth of the SMBH, its mass reaches a threshold level that triggers strong feedback, leading to gas ejection from the host galaxy.}
\label{fig:Galaxy_evolution}
\end{figure*}

%---------------- %%% Section 5 %%% ---------------------%

\section{Conclusion}\label{section5}
In this study, we have presented the simulations of merging galaxies with \texttt{GIZMO}. Our simulation contains the major mergers of gas-poor and gas-rich galaxies as well as minor mergers. We find that the dynamics of GMCs play an essential role in galaxy mergers. Massive molecular gas particles can more efficiently lose their kinetic energy than atomic-gas particles due to stronger dynamical friction, which causes them to fall into the galaxy's center. Consequently, SMBHs can grow from ${10^{7}} \rm\, M_\odot$ to $10^{9}\, \rm M_\odot$ within 300 Myr, providing a possible explanation for the creation of massive SMBH ($10^{9}\, \rm M_\odot$) in the early universe.\par
The SFRs in the galaxy mergers also become higher than those in the isolated case, and $\sim$ 50\% of the star formation occurs in the central region; the SFRs can even reach the starburst phase during a period of wet major merger, where more than 65\% of star formation occurs in the center of the galaxy. This result strengthens the idea that the dynamics difference between atomic and molecular gas particles leads to the merger-driven starburst. Similarly, in the merger of gas-poor galaxies, the SFR can be increased by a factor of 2 to 3 in the galactic center area.
Furthermore, we can grow the bulge of a gas-poor {disk} galaxy through minor mergers, and the size and mass of the bulge can increase from 0.92 kpc to 1.9 kpc and from $4.7\times 10^{10}\, \rm M_\odot$ to $7\times 10^{10}\, \rm M_\odot$, respectively. A galaxy with a prominent bulge is likely to experience many minor mergers during its evolution. Based on the simulation, we suggest that the Milky Way has experienced many minor mergers but no major mergers. We summarize the scenarios of galaxy mergers based on the simulation in Figure \ref{fig:Galaxy_evolution}.

This study sheds light on the rapid growth of SMBHs and the bulge evolution in interacting galaxies. Combining sophisticated models with the upcoming high-sensitivity surveys by James Webb Space Telescope {\it (JWST) } and the Vera C. Rubin Observatory {\it (LSST)}, we will be able to unveil the physics behind the growth of SMBHs and their host galaxies in the early universe.

%\section{acknowledgments}
%\acknowledgements
%\begin{acknowledgements}

\newpage
$$\rm Acknowledgement $$
{Authors are grateful to the anonymous referee for his/her constructive suggestions that significantly improve the quality of this paper.   We also thank Yen-Chen Pan and You-Hua Chu for their valuable comments. S.L. thanks Ching-Yao Tang for his encouragement and discussion. This work was supported by the National Science and Technology Council, Taiwan, under grant no. MOST 110-2112-M-001-068-MY3 and MOST 110-2112-M-008-021-MY3, as well as by the Academia Sinica, Taiwan, under a career development award, grant no. AS-CDA-111-M04. Our computational resources are provided by the National Energy Research Scientific Computing Center (NERSC), a U.S. Department of Energy Office of Science User Facility operated under Contract No. DE-AC02-05CH11231, and the TIARA Cluster at the Academia Sinica Institute of Astronomy and Astrophysics (ASIAA).} 

%\end{acknowledgements} 

%\begin{acknowledgments} 
%\acknowledgments

%\end{acknowledgments}

%\appendix
%put the appendex here
%log


%Pre mark
\bibliography{sample631}{}
\bibliographystyle{aasjournal}

%\begin{thebibliography}{}
%\expandafter\ifx\csname natexlab\endcsname\relax\def\natexlab#1{#1}\fi
%\providecommand{\url}[1]{\href{#1}{#1}}
%\providecommand{\dodoi}[1]{doi:~\href{http://doi.org/#1}{\nolinkurl{#1}}}
%\providecommand{\doeprint}[1]{\href{http://ascl.net/#1}{\nolinkurl{http://ascl.net/#1}}}
%\providecommand{\doarXiv}[1]{\href{https://arxiv.org/abs/#1}{\nolinkurl{https://arxiv.org/abs/#1}}}

%\end{thebibliography}



% End of file.

\end{document}